\RequirePackage{fixltx2e}
\documentclass[12pt,a4paper]{iopart}
\pdfoutput=1
\usepackage[unicode=true,bookmarksopen=true,colorlinks=true,urlcolor=blue,anchorcolor=blue,citecolor=blue,filecolor=blue,linkcolor=blue,menucolor=blue,pagecolor=blue,linktocpage=true,pdfa=true]{hyperref}
\renewcommand{\mailto}[1]{\href{mailto:#1}{\texttt{#1}}}

\makeatletter
\providecommand*\@nameundef[1]{\expandafter\let\csname #1\endcsname\@undefined}
\@nameundef{equation*}
\@nameundef{endequation*}
\makeatother

\usepackage{amsmath}
\usepackage{amssymb}
\usepackage{amsfonts}
\usepackage{mathtools}
\mathtoolsset{mathic}
\usepackage{lmodern}
\usepackage[T1]{fontenc}
\DeclareFontShape{OMX}{cmex}{m}{n}{
  <-7.5> cmex7
  <7.5-8.5> cmex8
  <8.5-9.5> cmex9
  <9.5-> cmex10
}{}
\SetSymbolFont{largesymbols}{normal}{OMX}{cmex}{m}{n}
\SetSymbolFont{largesymbols}{bold}  {OMX}{cmex}{m}{n}

\DeclareMathAlphabet{\mathsfi}{OT1}{cmss}{m}{sl}
\DeclareMathAlphabet{\mathbfi}{OML}{cmm}{b}{it}

\usepackage{bbm}

\let\originalleft\left
\let\originalright\right
\renewcommand{\left}{\mathopen{}\mathclose\bgroup\originalleft}
\renewcommand{\right}{\aftergroup\egroup\originalright}

\renewcommand{\vec}[1]{{\ifnum9<1#1\mathbf{#1}\else\ifcat\noexpand#1\relax\boldsymbol{#1}\else\mathbfi{#1}\fi\fi}}

\newcommand{\eqend}[1]{\,\mathrm{#1}}

\newcommand{\mathi}{\mathrm{i}}
\newcommand{\total}{\mathop{}\!\mathrm{d}}
\newcommand{\laplace}{\mathop{}\!\bigtriangleup}
\newcommand{\abs}[1]{{\left\lvert{#1}\right\rvert}}

\usepackage[numbers,sort&compress]{natbib}
\allowdisplaybreaks

\begin{document}

\title{Approaches to linear local gauge-invariant observables in inflationary cosmologies}

\author{Markus B. Fr{\"o}b${}^\dag$, Thomas-Paul Hack${}^\ddag$ and Igor Khavkine${}^\S$}
\address{${}^\dag$Department of Mathematics, University of York,\\\hspace{0.5em}Heslington, York, YO10 5DD, United Kingdom}
\address{${}^\ddag$Institut f{\"u}r Theoretische Physik, Universit{\"a}t Leipzig,\\\hspace{0.5em}Br{\"u}derstra{\ss}e 16, 04103 Leipzig, Germany}
\address{${}^\S$Institute of Mathematics, Czech Academy of Sciences,\\\hspace{0.5em}\v{Z}itn{\'a} 25, 115 67 Praha 1, Czech Republic}

\eads{${}^\dag$\mailto{mbf503@york.ac.uk}, ${}^\ddag$\mailto{thomas-paul.hack@itp.uni-leipzig.de}, ${}^\S$\mailto{khavkine@math.cas.cz}}

\begin{abstract}
We review and relate two recent complementary constructions of linear local gauge-invariant observables for cosmological perturbations in generic spatially flat single-field inflationary cosmologies. After briefly discussing their physical significance, we give explicit, covariant and mutually invertible transformations between the two sets of observables, thus resolving any doubts about their equivalence. In this way, we get a geometric interpretation and show the completeness of both sets of observables, while previously each of these properties was available only for one of them.
\end{abstract}

\submitto{CQG}


\section{Introduction}

Linear gauge-invariant observables play a crucial role in the theory of cosmological perturbations~\cite{mfb-pert}, since they obviously separate physical effects from gauge artifacts. Furthermore, \emph{local} gauge-invariant observables (those given by differential operators acting on the perturbations) are important because they separate the issue of gauge invariance from infrared effects, since gauge-invariant field combinations can be smeared by test functions with compact support. The usual gauge-invariant Bardeen potentials~\cite{bardeen,mfb-pert} are not local in this sense, because they are non-polynomial in spatial momenta. We elaborate on both of these aspects below, in Section~\ref{sec:context}. A set of local gauge-invariant observables is considered \emph{complete} if any field configuration for which they all vanish is (at least locally) constrained to be pure gauge. Consider any local tensor $T[g,\phi]$ covariantly constructed from the metric $g$ and (if any) matter fields $\phi$ that vanishes on a given background; by the well-known Stewart--Walker lemma~\cite[Lem.~2.2]{sw74} its linearisation $T^{(1)}$ is a local gauge-invariant observable, due to the Lie derivative identity
\begin{equation}
\mathcal{L}_v T[g,\phi] = T^{(1)}[\mathcal{L}_v (g,\phi)] \eqend{.}
\end{equation}
However, the question of the completeness of a set of local gauge-invariant observables obtained in this way has to be addressed separately.

Despite their potential importance, in the context of single-field inflationary cosmologies, such complete sets of local gauge-invariant observables were obtained only very recently. First, \cite{fhh17} gave such a set together with an explicit proof of its completeness. It was constructed by generalising from the previously studied cases of Minkowski and de~Sitter backgrounds~\cite{higuchi-ds}. However, the question of giving a Stewart--Walker type geometric interpretation to the explicit formulas of~\cite{fhh17} was left open. Shortly thereafter, the independent work~\cite{cdk17} addressed a related non-linear geometric problem (the characterisation of a particular isometry class of inflationary spacetimes by local tensorial equations). By the Stewart--Walker lemma, the linearisations of these tensor equations then give local gauge invariant observables. A heuristic argument given in~\cite{cdk17} implies that this set should also be complete. However,
the linearisation was not explicitly performed and the completion of the heuristic argument for completeness to a rigorous one was left open.

In this work, we rectify both of these problems by explicitly relating the observables of~\cite{fhh17} to the linearisations of the tensors of~\cite{cdk17} and vice versa. Thus, in one direction we propagate the proof of completeness, and in the other direction we propagate a geometric interpretation. In Sections~\ref{sec:cdk} and~\ref{sec:lin-fhh}, we respectively review the relevant results from~\cite{cdk17} and~\cite{fhh17}, reproducing the explicit formulas that we will need. In Section~\ref{sec:cdk-fhh}, the two sets of observables are related by explicit transformations. We conclude in Section~\ref{sec:discussion} with a discussion of our results and of directions for future work.

We assume that the spacetime dimension is $n \geq 3$. Also, we use the $({-}{+}\cdots{+})$ signature for the metric and more generally the sign conventions of~\cite{mtw} (which are coded as ``+++'' in their reference table). While we draw formulas and identities from both~\cite{fhh17} (the FHH paper) and~\cite{cdk17} (the CDK paper), we will mostly use the notational conventions of~\cite{fhh17}. They are related as follows:
\begin{table}[h]
\centering
\begin{tabular}{l|cccccccccc}
Notation of CDK & $m$ & $f(t)$ & $(-)'=\partial_t$ & $U^\mu$ & $\xi$ & $\vec{\eta}$ & $\mathcal{R}$ & $W_{ijkh}$ & $\kappa$ & $V(\phi)$ \\
\hline
\rule{0pt}{1.1em}Notation of FHH & $n-1$ & $a(\eta)$ & $(-)'=\partial_\eta$ & $u^\mu$ & $H$ & $\dot{H}$ & $R$ & $C_{\mu\nu\rho\sigma}$ & $\kappa^2/2$ & $V(\phi)$
\end{tabular}
\end{table}

\subsection{Cosmological context} \label{sec:context}

Before proceeding to the main body of this work we shall put our results into context and comment on the relevance of the fact that the observables we discuss are gauge-invariant, local and complete.

\paragraph{How do gauge invariant variables help with quantization?}
Let us denote the linearised perturbations of the metric and inflaton field by $h$ and $\psi$, respectively. Free field quantisation promotes them to quantum fields $\hat{h}$ and $\hat\psi$, with canonical commutation relations $[ \hat{h}(x), \hat{h}(y) ] = \mathi \hbar P_{hh}(x,y)$ and so on, fixed by the classical Poisson brackets $P_{hh}(x,y) = \{ h(x), h(y) \}$. As is well known, in gauge theories the Poisson brackets $P(x,y)$ strongly depend on the choice of gauge fixing, the choice of primary quantisation variables, etc. However, as is also well known, none of these choices affect local gauge-invariant observables. For instance, if $A[h,\psi]$ and $B[h,\psi]$ are two local gauge-invariant observables linear in the fields, which we can smear to $A(\alpha) = \int \alpha(x) A(x) \total x$ and $B(\beta) = \int \beta(x) B(x) \total x$ by compactly supported test functions, then the value of
\begin{equation}
\{ A(\alpha), B(\beta) \} = \int A^*[\alpha](x) P(x,y) B^*[\beta](y) \total x \total y \eqend{,}
\end{equation}
where $A^*$ and $B^*$ are the adjoints of $A$ and $B$, is guaranteed to be independent of all of the above choices.

One pathology that occurs frequently is that $P(x,y) \neq 0$ when $x$ and $y$ are spacelike separated. Such non-causal canonical commutation relations are a hallmark of certain gauge fixing schemes, of which the Coulomb gauge in electrodynamics is a prime example. On the other hand, it is easy to show that causality is restored for gauge invariant observables. For instance, $\{ A(\alpha), B(\beta) \} = 0$ whenever the supports of $\alpha$ and $\beta$ are spacelike separated. One way to see this is to note that some gauge fixing and quantisation schemes, for example using harmonic gauge (cf.~\cite{Fewster:2012bj} for linearised gravity, or~\cite{Hack:2012dm,Khavkine:2014kya} for a much larger class of gauge theories), result in a $P(x,y)$ that is manifestly causal. But since $\{ A(\alpha), B(\beta) \}$ is independent of the gauge fixing, if it vanishes in harmonic gauge, it vanishes for all other gauge fixing schemes. Since harmonic gauge is not necessarily the most convenient gauge choice in cosmology, this is a comforting observation.

Another kind of pathology that may occur is the unexpected growth of a vacuum correlation function like $\langle \hat{h}(x) \hat{h}(y) \rangle$ at large separations between $x$ and $y$. A priori, it may not be clear whether such behaviour is indeed a physical effect, or needs to be ascribed to a gauge artifact, a poor choice of renormalisation scheme, or an unphysical choice of the vacuum state. In this case, examining instead the correlation functions $\langle A[\hat{h}](x) B[\hat{h}](y) \rangle$, where $A$ and $B$ range through a complete set of local gauge-invariant observables, can discriminate the case of a gauge artifact from all other possibilities. Such asymptotic growth of correlators has been the subject of quite some controversy in the recent cosmological literature~\cite{tsamis_woodard_npb_1996,tsamis_woodard_ap_1997,garriga_tanaka_prd_2008,tsamis_woodard_prd_2008,janssen_prokopec_cqg_2008,riotto_sloth_jcap_2008,burgess_et_al_jcap_2010,rajaraman_kumar_leblond_prd_2010,seery_cqg_2010,urakawa_tanaka_prd_2010,kahya_onemli_woodard_plb_2011,giddings_sloth_jcap_2011,gerstenlauer_et_al_jcap_2011,giddings_sloth_prd_2011,garbrecht_rigopoulos_prd_2011,xue_gao_brandenberger_jcap_2012,pimentel_senatore_zaldarriaga_jhep_2012,senatore_zaldarriaga_jhep_2013,assassi_baumann_green_jhep_2013,urakawa_tanaka_cqg_2013,tanaka_urakawa_ptep_2014}, and an analysis through the lens of the complete sets of local gauge-invariant observables described in this work could help to resolve the issue, see also~\cite{fhh17,Higuchi:2011vw,Higuchi:2017sgj}.

\paragraph{How non-local are Bardeen potentials?}
The standard set of gauge-invariant field combinations for cosmological perturbations are the so-called Bardeen potentials~\cite{bardeen,mfb-pert}. In $n = 4$ dimensions, they are usually constructed from the following parametrisation of the perturbed metric:
\begin{equation}
g^{(1)} = a(\eta)^2 \left[ - 2 A \total\eta^2 + 2 \left( \partial_i B + \hat{B}_i \right) \total x^i \total\eta + h_{ij} \total x^i \total x^j \right] \eqend{,}
\end{equation}
where
\begin{equation}
h_{ij} = 2 C \delta_{ij} + 2 \left( \partial_i \partial_j - \frac{1}{3} \delta_{ij} \laplace \right) E + 2 \partial_{(i} \hat{E}_{j)} + \hat{E}_{ij} \eqend{,}
\end{equation}
with $\partial^i \hat{B}_i = 0$, $\partial^i \hat{E}_i = 0$, $\partial^i \hat{E}_{ij} = 0$ and $\hat{E}_i^i = 0$, the spatial indices are raised and lowered with $\delta_{ij}$ and $\laplace = \partial^i \partial_i$. Besides, $\hat{E}_{ij}$, the gauge-invariant Bardeen potentials $\{ \Psi, \Phi, \hat{\Phi}_i, \hat{E}_{ij} \}$ consist of
\begin{subequations}
\begin{align}
\Psi &= A + a H (B-E') + (B-E')' \eqend{,} \\
\Phi &= - C - a H (B-E') + \frac{1}{3} \laplace E \eqend{,} \\
\hat{\Phi}_i &= \hat{E}'_i - \hat{B}_i \eqend{,}
\end{align}
\end{subequations}
where a prime denotes a derivative with respect to conformal time $\eta$, and $H = a'/a^2$ is the Hubble rate. However, to explicitly write the gauge-invariant fields in terms of the components of $A$, $B_i = \partial_i B + \hat{B}_i$ and $h_{ij}$, we must use non-local
expressions; for example
\begin{equation}
\hat{E}_{ij} = h_{ij} - \frac{1}{2} \left( \delta_{ij} - \frac{\partial_i \partial_j}{\laplace} \right) h_k^k + \frac{1}{2} \left( \delta_{ij} + \frac{\partial_i \partial_j}{\laplace} \right) \frac{\partial^k \partial^l}{\laplace} h_{kl} - 2 \frac{\partial^k}{\laplace} \partial_{(i} h_{j)k} \eqend{.}
\end{equation}
Each application of the inverse spatial Laplacian $\laplace^{-1}$ corresponds to convolution with respect to the usual Newtonian potential $\abs{\vec{x}-\vec{x}'}^{-1}$. Thus, even though $\hat{E}_{ij}$ is gauge invariant, its expression in terms of $h_{ij}$ is ultimately non-local and has the form
\begin{equation}
\hat{E}_{ij}(t,\vec{x}) = \int E_{ij}^{kl}(\vec{x},\vec{x}') h_{kl}(t,\vec{x}') \total^3 x' \eqend{,}
\end{equation}
where the integral kernel behaves as $E_{ij}^{kl}(\vec{x},\vec{x}') \sim \abs{\vec{x}-\vec{x}'}^{-3}$, for large comoving spacelike distances, $\abs{\vec{x}-\vec{x}'} \to \infty$.

This non-locality of the Bardeen potentials has various physical consequences. We have already mentioned the issue of non-vanishing commutators at spacelike separations as a possible gauge-artifact, though it can also occur for manifestly gauge-invariant expressions. For example, the commutators $[ \hat\Psi(x), \hat\Psi(y) ]$ and $[ \hat\Phi(x), \hat\Phi(y) ]$ of the quantised Bardeen potentials $\hat\Psi$ and $\hat\Phi$ do not vanish at spacelike separated points~\cite{Eltzner:2013soa,Hack:2014epa}, whereas this problem does not occur for \emph{local} gauge-invariant observables~\cite{Fewster:2012bj,Hack:2014epa}.

\paragraph{Can our constructions be extended to anisotropic cosmologies?}
It is well known that a complete set of local gauge-invariant observables on Minkowski spacetime is given by the components of the linearised Riemann tensor. For constant curvature spacetimes (like de~Sitter or anti-de~Sitter), the components of the linearised Riemann tensor also do the job, provided its tensor indices are appropriately raised or lowered before linearisation~\cite{Frob:2014cza,khavkine_jgeomphys_2014}. Unfortunately, no simple generalisation of these results exists for an arbitrary spacetime, which explains the necessity of the works~\cite{fhh17,cdk17} and their present conclusion to produce the answer for spatially flat inflationary cosmologies. Any other spacetime, or family of spacetimes, like in the Bianchi homogeneous anisotropic cosmology classification, needs to be analysed separately.

The approach of~\cite{cdk17} outlines an effective strategy for producing a candidate for a complete set of local gauge-invariant observables, together with their Stewart--Walker type geometric interpretations: the IDEAL characterisation tensors (cf.~Section~\ref{sec:cdk}). Suppose that a given spacetime (a) can be locally characterised by the existence of certain tensor fields $X$, $Y$, $Z$, \ldots satisfying some covariant differential conditions $C(X,\nabla X, Y, \ldots)$, $D(X,\nabla X, Y, \ldots)$,
\ldots\ and (b) each of these tensor fields may be covariantly constructed from the dynamical fields of the theory: the metric, curvature tensor and its derivatives, $X = X(\phi,\nabla \phi,g,R,\nabla R,\ldots)$, $Y = (\phi,\nabla \phi,g,R,\nabla R, \ldots)$, \ldots.\ Then the list of tensors obtained by plugging the expressions from (b) into those
from (a), gives a list of IDEAL characterisation tensors $C(X(\phi,\nabla \phi,g,R,\ldots), \nabla X(\phi,\nabla \phi,g,R,\ldots), \ldots)$, \ldots, whose linearisations are a good candidate for a complete set of local gauge-invariant observables on the given spacetime. The success of this strategy for homogeneous and isotropic cosmologies in~\cite{cdk17} relied greatly on type (a) results already existing in the literature. Previous IDEAL characterisations of other spacetimes~\cite{0264-9381-15-5-014,Ferrando:2002dq,Ferrando:2008nw,Ferrando:2010ni,Ferrando:2017idk} have also succeeded for similar reasons. We have not investigated the existing literature for type (a) results for other Bianchi cosmological families. Strictly speaking, any such candidate set of observables should also be rigorously checked for completeness, as was done in~\cite{fhh17}. However, it is not clear to what other spacetime families the methods of~\cite{fhh17} can be generalised; a promising systematic approach to this problem is currently being investigated~\cite{aabk-private}.

\paragraph{Can our constructions be extended to higher perturbative orders?}
Our constructions may not be directly generalised to higher perturbative orders. In fact, as is well-known, strictly local gauge-invariant observables, i.e., gauge-invariant observables which are of the form $O(x) = P(g(x),\nabla g(x),\dots,\phi(x),\nabla \phi(x),\dots)$, with $P$ a polynomial, do not exist in diffeomorphism-invariant theories. Instead, one may consider so-called relational observables --- see, e.g.,~\cite{Giddings:2005id,Brunetti:2013maa,kh-locobsv,Marolf:2015jha} and~\cite{Tambornino:2011vg} for a recent review containing further references --- which characterise the state of one dynamical field with respect to another dynamical field, and are mildly non-local. These have been used in the cosmological context in~\cite{Brunetti:2016hgw,Frob:2017gyj} and have a clear geometric interpretation. However, an explicit proof of their completeness is not available yet.

\paragraph{Other physical applications.} In addition to the benefits of being local and having the improved physical properties which follow from this, the observables discussed in this work are manifestly complete. This implies that the knowledge of their correlation functions is sufficient to discriminate between different quantum states. An analysis, based for instance on the framework of~\cite{Fewster:2012bj}, of detailed consequences and applications of this fact would be interesting, but is beyond the scope of this work.

\section{CDK tensors}
\label{sec:cdk}

An IDEAL characterisation of generic spatially flat single-field inflationary spacetime was given in~\cite{cdk17}. It consists of a set of tensors that vanish on a given spacetime and scalar field $(M,g,\phi)$ if and only if it is locally isometric to a particular isometry class of an FLRW cosmology $(M,g)$ with an inflaton field $\phi$ with potential $V(\phi)$ (up to some non-degeneracy conditions). The particular isometry class is identified by a scalar function $\Xi(\phi)$ satisfying the auxiliary (``Hamilton--Jacobi''~\cite{salopek-bond}) equation
\begin{equation}
\left[ \Xi'(\phi) \right]^2 - \frac{n-1}{2 (n-2)} \kappa^2 \Xi(\phi)^2 + \frac{\kappa^4}{4 (n-2)^2} V(\phi) = 0 \eqend{.}
\end{equation}
We define
\begin{equation}
u_\mu \equiv \frac{\nabla_\mu \phi}{\sqrt{ - \nabla^\rho \phi \nabla_\rho \phi}} \eqend{,} \qquad H \equiv \frac{\nabla^\mu u_\mu}{n-1} \eqend{,}
\end{equation}
and denote with an overdot the derivative in the direction of $u^\mu$, i.\,e. $\dot H \equiv u^\mu \nabla_\mu H$, $\dot \phi \equiv u^\mu \nabla_\mu \phi = - \sqrt{ - \nabla^\mu \phi \nabla_\mu \phi}$, etc. The IDEAL characterisation of~\cite{cdk17} consists of the tensors
\begin{subequations}
\label{eq:cdk-def}
\begin{align}
\mathfrak{Z}_{\mu\nu\rho\sigma} &\equiv R_{\mu\nu\rho\sigma} - 2 H^2 g_{\mu[\rho} g_{\sigma]\nu} - 4 \dot{H} u_{[\mu} g_{\nu][\rho} u_{\sigma]} \eqend{,} \\
\mathfrak{D}_{\mu\nu} &\equiv \nabla_\mu u_\nu - \left( g_{\mu\nu} + u_\mu u_\nu \right) H \eqend{,} \\
\mathfrak{P}_{\mu\nu} &\equiv u_{[\mu} \nabla_{\nu]} H \eqend{,} \\
\mathfrak{S} &\equiv \dot\phi + \frac{2 (n-2)}{\kappa^2} \Xi'(\phi) \eqend{,} \\
\mathfrak{T} &\equiv H - \Xi(\phi) \eqend{.}
\end{align}
\end{subequations}
We call these the \emph{CDK tensors}. Physically, $u^\mu$ is the future-pointing normal vector to slices of constant cosmological time, $H$ corresponds to the Hubble rate and $\dot H$ to the Hubble acceleration with respect to cosmological proper time. Note that these interpretations are valid only if \emph{all} of the CDK tensors vanish, which is what guarantees that the underlying geometry is of an inflationary cosmology. Also, it is important to mention that these equations are applicable when the following non-degeneracy conditions are satisfied on the entire spacetime:
\begin{equation}
\nabla^\mu \phi \nabla_\mu \phi < 0 \eqend{,} \qquad \Xi(\phi) \neq 0 \eqend{,} \qquad \Xi'(\phi) > 0 \eqend{,} \qquad \text{and} \quad V'(\phi) \neq 0 \eqend{.}
\end{equation}
In a special case where some of these conditions are not satisfied, a slightly different set of equations can still provide an IDEAL characterisation. A list of \emph{regular inflationary cosmologies}, where this is possible, and a complete classification of corresponding IDEAL characterisations was given in~\cite{cdk17}, including cases with positive or negative spatial curvature and with a massless inflaton.

An IDEAL characterisation is necessarily far from unique. Once one has been given, many others can be produced from it by taking algebraic and differential expressions of the original set of tensors, as long as these expressions can be inverted on an open neighborhood\footnote{This statement may be interpreted in the context of functional spaces with the Whitney strong topology, or alternatively in the usual manifold topology of jet bundles of sufficiently high order. See the discussion of topological issues in~\cite{kh-locobsv} for more details.} of the characterised isometry class. For instance, note the identities
\begin{subequations}
\begin{align}
\mathfrak{P}_{\mu\nu} &= u_{[\mu} \nabla_{\nu]} \mathfrak{T} \eqend{,} \\
\mathfrak{D}_{[\mu\nu]} &= \nabla_{[\mu} u_{\nu]} = \left( \dot\phi \right)^{-1} u_{[\mu} \nabla_{\nu]} \mathfrak{S} \eqend{,} \\
u^\mu u^\nu \mathfrak{D}_{\mu\nu} &= u^\mu u^\nu \nabla_\mu u_\nu = \frac{1}{2} u^\mu \nabla_\mu (u^2) = 0 \eqend{,} \\
g^{\mu\nu} \mathfrak{D}_{\mu\nu} &= 0 \eqend{,}
\end{align}
\end{subequations}
the first two of which rely on $u_\mu$ being proportional to $\nabla_\mu \phi$. The logic behind the precise choice of the CDK tensors is explained in~\cite{cdk17}. For our purposes, it is more convenient to replace these tensors by the slightly different set $\{ W_{\mu\nu\rho\sigma}, \hat{\mathfrak{Z}}_{\mu\nu}, \hat{\mathfrak{D}}_{\mu\nu}, \mathfrak{S}, \mathfrak{T} \}$, where the first three tensors are defined as
\begin{subequations}
\label{eq:cdk2-def}
\begin{align}
W_{\mu\nu\rho\sigma} &\equiv \mathfrak{Z}_{\mu\nu\rho\sigma} - \frac{2}{n-2} \left( g_{\rho[\mu} \mathfrak{Z}_{\nu]\sigma} - g_{\sigma[\mu} \mathfrak{Z}_{\nu]\rho} \right) + \frac{2}{(n-2) (n-1)} g_{\mu[\rho} g_{\sigma]\nu} \mathfrak{Z} \eqend{,} \\
\hat{\mathfrak{Z}}_{\mu\nu} &\equiv \mathfrak{Z}_{\mu\nu} - \frac{1}{2} \mathfrak{Z} g_{\mu\nu} \eqend{,} \\
\hat{\mathfrak{D}}_{\mu\nu} &= \mathfrak{D}_{(\mu\nu)} - 2 u_{(\mu} \delta_{\nu)}^\rho u^\sigma \mathfrak{D}_{(\rho\sigma)} \eqend{,}
\end{align}
\end{subequations}
and where we have used the notation $\mathfrak{Z}_{\mu\nu} = g^{\rho\sigma} \mathfrak{Z}_{\mu\rho\nu\sigma}$, $\mathfrak{Z} = g^{\mu\nu}\mathfrak{Z}_{\mu\nu}$ and $\hat{\mathfrak{Z}} = g^{\mu\nu}\hat{\mathfrak{Z}}_{\mu\nu}$. We will still refer to them as CDK tensors, since one can straightforwardly invert the definition:
\begin{subequations}
\begin{align}
\mathfrak{Z}_{\mu\nu\rho\sigma} &= W_{\mu\nu\rho\sigma} + \frac{2}{n-2} \left( g_{\rho[\mu} \hat{\mathfrak{Z}}_{\nu]\sigma} - g_{\sigma[\mu} \hat{\mathfrak{Z}}_{\nu]\rho} \right) - \frac{4}{(n-2) (n-1)} g_{\mu[\rho} g_{\sigma]\nu} \hat{\mathfrak{Z}} \eqend{,} \\
\mathfrak{D}_{(\mu\nu)} &= \hat{\mathfrak{D}}_{\mu\nu} + u^\rho u_{(\mu} \hat{\mathfrak{D}}_{\nu)\rho} \eqend{.}
\end{align}
\end{subequations}
One checks by inserting the definition of $\mathfrak{Z}_{\mu\nu\rho\sigma}$ that $W_{\mu\nu\rho\sigma}$ is none other than the Weyl tensor $C_{\mu\nu\rho\sigma}$. Also, it is still true that $u^\mu u^\nu \hat{\mathfrak{D}}_{\mu\nu} = 0 = g^{\mu\nu} \hat{\mathfrak{D}}_{\mu\nu}$.

\section{Linearised FHH tensors}
\label{sec:lin-fhh}

The authors of~\cite{fhh17} gave a set of linear differential operators acting on the linearised metric and scalar field perturbation on a generic spatially flat inflationary spacetime $(M,g,\phi)$, together with explicit proofs of their invariance under linearised gauge transformations (diffeomorphisms) and their completeness (meaning that a field annihilated by all of these operators must locally be a pure gauge mode). This set was constructed by generalising similar known results for Minkowski and de~Sitter spacetimes~\cite{higuchi-ds}. We refer to this set as the \emph{linearised FHH tensors}.

Consider the background metric $g_{\mu\nu} = a^2 \, \eta_{\mu\nu}$ given by a conformal rescaling of the Minkowski metric $\eta_{\mu\nu}$ by the \emph{scale factor} $a = a(\eta)$ in coordinates $(x^\mu) = (x^0=\eta,x^i)$, together with a background scalar field $\phi = \phi(\eta)$. Denoting a derivative with respect to conformal time $\eta$ with a prime, we define the Hubble and slow-roll parameters
\begin{equation}
\label{H_epsilon_delta_def}
H \equiv \frac{a'}{a^2} \eqend{,} \qquad \epsilon \equiv - \frac{H'}{H^2 a} \eqend{,} \qquad \delta \equiv \frac{\epsilon'}{2 H a \epsilon} \eqend{.}
\end{equation}
An inflationary geometry satisfies the Einstein--Klein--Gordon equations with a potential $V(\phi)$, namely
\begin{subequations}
\begin{align}
E_{\mu\nu} &\equiv 2 R_{\mu\nu} - R g_{\mu\nu} - \kappa^2 \left[ \nabla_\mu \phi \nabla_\nu \phi - \frac{1}{2} g_{\mu\nu} \left( \nabla^\rho \phi \nabla_\rho \phi + V(\phi) \right) \right] = 0 \eqend{,} \\
F &\equiv \nabla^\mu \nabla_\mu \phi - \frac{1}{2} V'(\phi) = 0 \eqend{.} \label{scalar_eom}
\end{align}
\end{subequations}
As is well-known, as long as $\phi$ is not constant (or more precisely has a non-null gradient), the Klein--Gordon equation follows as an integrability condition of the Einstein equations, by the identity
\begin{equation}
\label{F_from_Emunu}
F = - \frac{\nabla^\mu \phi \nabla^\nu E_{\mu\nu}}{\kappa^2 \nabla^\rho \phi \nabla_\rho \phi} \eqend{.}
\end{equation}
Taking into account the assumed form of $a = a(\eta)$ and $\phi = \phi(\eta)$, these equations reduce to the well-known Friedmann equations for the background:
\begin{subequations}
\label{background_phiv}
\begin{align}
\kappa^2 V(\phi) &= 2 (n-2) (n-1-\epsilon) H^2 \eqend{,} \\
\kappa^2 (\phi')^2 &= 2 (n-2) \epsilon H^2 a^2 \eqend{,} \\
\phi'' &= - (n-2) H a \phi' - \frac{1}{2} a^2 V'(\phi) = (1-\epsilon+\delta) H a \phi' \eqend{.}
\end{align}
\end{subequations}

We parametrise the linear perturbations in $g_{\mu\nu} + g_{\mu\nu}^{(1)} + \cdots$ and $\phi + \phi^{(1)} + \cdots$ as
\begin{equation}
g_{\mu\nu}^{(1)} = a^2 h_{\mu\nu} \eqend{,} \qquad \phi^{(1)} = \frac{\phi'}{H a} \psi \eqend{.}
\end{equation}
Under linearised diffeomorphisms with parameter $\xi_\mu$, they transform as
\begin{equation}
\delta_\xi h_{\mu\nu} = 2 \partial_{(\mu} \xi_{\nu)} - 2 H a \eta_{\mu\nu} \xi_0 \eqend{,} \qquad \delta_\xi \psi = - H a \xi_0 \eqend{.}
\end{equation}
The explicit form of the linearised FHH tensors is as follows, where the derivative operators $\partial_\mu$ and the tensor components are all with respect to the $(x^\mu)$ coordinates, and where $\phi' \neq 0$ in the whole spacetime was assumed as a non-degeneracy condition:
\begin{subequations}
\label{eq:lin-fhh-def}
\begin{align}
\begin{split}
C_{\mu\nu\rho\sigma}^{(1)} &= a^2 \bigg[ \partial_\nu \partial_{[\rho} h_{\sigma]\mu} - \partial_\mu \partial_{[\rho} h_{\sigma]\nu} - \frac{4}{n-2} \left( \eta_{\rho[\mu} R^\text{flat}_{\nu]\sigma} - \eta_{\sigma[\mu} R^\text{flat}_{\nu]\rho} \right) \\
&\qquad\quad+ \frac{2}{(n-2) (n-1)} \eta_{\mu[\rho} \eta_{\sigma]\nu} R^\text{flat} \bigg] \eqend{,}
\end{split} \\
\begin{split}
E_{\mu\nu}^{(1)} &= 2 R^\text{flat}_{\mu\nu} - \eta_{\mu\nu} R^\text{flat} - (n-2) H a \left( 2 \partial_{(\mu} h_{\nu)0} - h_{\mu\nu}' + 4 \epsilon \delta^0_{(\mu} \partial_{\nu)} \psi \right) \\
&\quad+ (n-2) H a \eta_{\mu\nu} \left( 2 \partial^\rho h_{0\rho} - h' - 2 \epsilon \psi' \right) - 4 (n-2) \epsilon \delta H^2 a^2 \left( \eta_{\mu\nu} + \delta^0_\mu \delta^0_\nu \right) \psi \\
&\quad- (n-2) (n-1-\epsilon) H^2 a^2 \eta_{\mu\nu} \left( h_{00} + 2 \epsilon \psi \right) \eqend{,}
\end{split} \\
C_{\mu\nu}^{(1)} &= \frac{2}{n-2} R^\text{flat}_{\mu\nu} - \frac{1}{(n-1) (n-2)} \eta_{\mu\nu} R^\text{flat} + 2 \partial_\mu \partial_\nu \psi \eqend{,}
\end{align}
\end{subequations}
where
\begin{equation}
R^\text{flat}_{\mu\nu} = \partial_{(\mu} \partial^\rho h_{\nu)\rho} - \frac{1}{2} \partial^2 h_{\mu\nu} - \frac{1}{2} \partial_\mu \partial_\nu h \eqend{,} \qquad R^\text{flat} = \eta^{\mu\nu} R^\text{flat}_{\mu\nu} = \partial^\mu \partial^\nu h_{\mu\nu} - \partial^2 h
\end{equation}
are the linearised flat-space Ricci tensor and scalar. By the notation, we mean that $C_{\mu\nu\rho\sigma}^{(1)}$ and $E_{\mu\nu}^{(1)}$ are the linearisations of the Weyl tensor and the Einstein equations, respectively. To avoid confusion, we note that $C_{\mu\nu}^{(1)} \not\equiv g^{\rho\sigma} C_{\mu\rho\nu\sigma}^{(1)} = 0$, with the latter equality as expected for the linearised Weyl tensor. We have not yet defined a tensor $C_{\mu\nu}$ whose linearisation coincides with the formula for $C_{\mu\nu}^{(1)}$. Such a tensor will actually be defined in the next section, solving the problem (left open in~\cite{fhh17}) of finding a geometric interpretation for all of the components of $C_{\mu\nu}^{(1)}$. It is again convenient to choose a slightly different tensor, taking
\begin{equation}
\begin{split}
\hat{C}_{\mu\nu}^{(1)} &\equiv C_{\mu\nu}^{(1)} - \frac{1}{n-2} E_{\mu\nu}^{(1)} + \frac{1}{(n-1) (n-2)} E^{(1)} g_{\mu\nu} + u_\mu u_\nu \frac{2 H}{\dot \phi} F^{(1)} \\
&= 2 \partial_\mu \partial_\nu \psi + 2 \delta^0_\mu \delta^0_\nu \partial^2 \psi + H a \left( 2 \partial_{(\mu} h_{\nu)0} - h_{\mu\nu}' + 4 \epsilon \delta^0_{(\mu} \partial_{\nu)} \psi + \frac{2}{n-1} \eta_{\mu\nu} \epsilon \psi' \right) \\
&\quad+ H a \delta^0_\mu \delta^0_\nu \left[ 2 \partial^\rho h_{0\rho} - h' - 2 (n-2+2\delta) \psi' \right] - 2 \delta^0_\mu \delta^0_\nu H^2 a^2 \delta \left( h_{00} + 2 \epsilon \psi \right) \\
&\quad- (n-1-\epsilon) H^2 a^2 \left( \frac{\eta_{\mu\nu}}{n-1} + 2 \delta^0_\mu \delta^0_\nu \right) \left( h_{00} + 2 \epsilon \psi \right) \eqend{,}
\end{split}
\end{equation}
where $E^{(1)} = g^{\mu\nu} E_{\mu\nu}^{(1)}$, and $F^{(1)}$ is the linearisation of the Klein--Gordon equation~\eqref{scalar_eom}. Since the formula for $C_{\mu\nu}^{(1)}$ of~\cite{fhh17} was anyway only given for on-shell fields (those annihilated by $E_{\mu\nu}^{(1)}[h,\psi]$ and $F^{(1)}[h,\psi]$), these additional terms can be simply seen as our preferred choice of an off-shell representative.

\section{Relating CDK and FHH tensors}
\label{sec:cdk-fhh}

The well-known Stewart--Walker lemma~\cite[Lem. 2.2]{sw74} (slightly generalised to include scalar fields) states that the linearisation $T^{(1)}[h,\psi]$ of a tensor $T[g,\phi]$ locally and covariantly constructed out of the metric $g$, the scalar field $\phi$ and their derivatives is gauge-invariant when linearised around a background where $T[g,\phi] = 0$. All the tensors from an IDEAL characterisation of a particular isometry class satisfy the hypotheses of the Stewart--Walker lemma, when linearised on a background belonging to this isometry class, and hence give gauge-invariant linear observables depending locally on $h$ and $\psi$. In the Introduction of~\cite{cdk17}, it was also argued that the defining property of an IDEAL characterisation (that the joint kernel of these tensors is locally exhausted by representatives of the given isometry class) generically implies that the linearisation of these tensors gives a complete set of gauge-invariant observables (meaning that a field for which all of them vanish must locally be a pure gauge mode) on a single-field inflationary spacetime. However, in~\cite{cdk17}, the linearisations were not explicitly computed, nor was their completeness explicitly checked.

In~\cite{fhh17}, explicit proofs were given for the gauge invariance and completeness of a set of linear local observables on a generic single-field inflationary spacetime. However, the geometric interpretations of these observables remained partially obscure: while some of the observables were found to come from the linearisation of covariantly constructed tensors vanishing on the
background, as in the Stewart--Walker lemma, not all of them could be interpreted in this way. Below, we rectify both of the above mentioned deficiencies by relating the two constructions to each other. Our strategy is to identify a set of tensors $\{ C_{\mu\nu\rho\sigma}, E_{\mu\nu}, \hat{C}_{\mu\nu} \}$ covariantly constructed from the metric $g$ and the inflaton scalar $\phi$, which we call the \emph{FHH tensors}, that
\begin{itemize}
\item constitute an IDEAL characterisation of a generic spatially flat inflationary spacetime, by expressing them in terms of the CDK tensors~\eqref{eq:cdk-def}, \eqref{eq:cdk2-def} in an invertible way, and
\item reproduce the linear operators from Section~\ref{sec:lin-fhh} upon linearisation.
\end{itemize}
This would already provide a complete geometric interpretation for the linear observables constructed in~\cite{fhh17}. Then the linearisations of the relations between the CDK and FHH tensors will give an explicit and invertible relationship between the linearised CDK and FHH tensors. This, in turn, would show that the linearised CDK tensors do in fact give a complete set of gauge-invariant linear observables.

In terms of the CDK tensors, the FHH tensors have the following expressions:
\begin{subequations}
\label{eq:fhh-from-cdk}
\begin{align}
C_{\mu\nu\rho\sigma} &= W_{\mu\nu\rho\sigma} \eqend{,} \\
\begin{split}
E_{\mu\nu} &= 2 \hat{\mathfrak{Z}}_{\mu\nu} - g_{\mu\nu} \left[ \frac{\kappa^2}{2} \mathfrak{S}^2 + (n-1) (n-2) \left( 2 H - \mathfrak{T} \right) \mathfrak{T} \right] \\
&\quad- u_\mu u_\nu \kappa^2 \dot\phi \mathfrak{S} - 2 (n-2) \left( g_{\mu\nu} + u_\mu u_\nu \right) \dot{\mathfrak{T}} \eqend{,}
\end{split} \\
\begin{split}
\hat{C}_{\mu\nu} &= - 2 H \hat{\mathfrak{D}}_{\mu\nu} - 2 H \left( g_{\mu\nu} + n u_\mu u_\nu \right) \mathfrak{T} \\
&\quad+ g_{\mu\nu} \frac{\kappa^2 \dot\phi}{(n-1) (n-2)} \mathfrak{S} + 2 H u_\mu u_\nu \left( H n + \frac{V'(\phi)}{\dot\phi} \right) \frac{\mathfrak{S}}{\dot\phi} \eqend{.}
\end{split}
\end{align}
\end{subequations}
The validity of these formulas can be checked by direct calculation. The formula for $\hat{C}_{\mu\nu}$ serves as its definition. The agreement of its linearisation with $\hat{C}^{(1)}_{\mu\nu}$~\eqref{eq:lin-fhh-def} can also be checked by direct calculation. We now reverse the direction and, by straightforward algebraic manipulations, express the CDK tensors in terms of the FHH tensors:
\begin{subequations}
\label{eq:cdk-from-fhh}
\begin{align}
\hat{\mathfrak{D}}_{\mu\nu} &= - \frac{1}{2 H} \left[ \hat{C}_{\mu\nu} - u_\mu u_\nu \left( u^\rho u^\sigma \hat{C}_{\rho\sigma} \right) - \frac{g_{\mu\nu} + u_\mu u_\nu}{n-1} \left( g^{\rho\sigma} + u^\rho u^\sigma \right) \hat{C}_{\rho\sigma} \right] \eqend{,} \\
\mathfrak{S} &= \frac{(n-1) (n-2) \dot\phi \, g^{\mu\nu} \hat{C}_{\mu\nu}}{n \kappa^2 \dot\phi^2 - 2 (n-1) (n-2) H \left( H n + V'(\phi)/\dot\phi \right)} \eqend{,} \\
\mathfrak{T} &= - \frac{1}{2 H (n-1)} \left( g^{\mu\nu} + u^\mu u^\nu \right) \hat{C}_{\mu\nu} + \frac{\kappa^2 \dot\phi}{2 H (n-1) (n-2)} \mathfrak{S} \eqend{,} \\
\begin{split}
\hat{\mathfrak{Z}}_{\mu\nu} &= \frac{1}{2} E_{\mu\nu} + \frac{1}{2} g_{\mu\nu} \left[ \frac{\kappa^2}{2} \mathfrak{S}^2 + (n-1) (n-2) \left( 2 H - \mathfrak{T} \right) \mathfrak{T} \right] \\
&\quad+ \frac{1}{2} u_\mu u_\nu \kappa^2 \dot\phi \mathfrak{S} + (n-2) \left( g_{\mu\nu} + u_\mu u_\nu \right) \dot{\mathfrak{T}} \eqend{,}
\end{split} \\
W_{\mu\nu\rho\sigma} &= C_{\mu\nu\rho\sigma} \eqend{.}
\end{align}
\end{subequations}
Note that $\mathfrak{S}$ is expressed only in terms of $\hat{C}_{\mu\nu}$. This formula then enters the expression for $\mathfrak{T}$. Both of these formulas then also enter the formula for $\hat{\mathfrak{Z}}_{\mu\nu}$. These expressions were not expanded fully for economy of notation.

When it comes to linearisation, either for writing the CDK tensors in terms of the FHH ones, or vice versa, the above formulas give the answer almost immediately. It suffices to replace the tensors from each set by their linearisation, ignore any terms of of quadratic order in the tensors from either set and, if helpful, replace remaining coefficients by their background values~\eqref{background_phiv}. The linearised FHH tensors in terms of the linearised CDK tensors are
\begin{subequations}
\label{eq:lin-fhh-from-cdk}
\begin{align}
C_{\mu\nu\rho\sigma}^{(1)} &= W_{\mu\nu\rho\sigma}^{(1)} \eqend{,} \\
\begin{split}
E_{\mu\nu}^{(1)} &= 2 \hat{\mathfrak{Z}}^{(1)}_{\mu\nu} - 2 (n-1) (n-2) H g_{\mu\nu} \mathfrak{T}^{(1)} \\
&\quad- u_\mu u_\nu \kappa^2 \dot\phi \mathfrak{S}^{(1)} - 2 (n-2) \left( g_{\mu\nu} + u_\mu u_\nu \right) u^\rho \nabla_\rho \mathfrak{T}^{(1)} \eqend{,}
\end{split} \\
\begin{split}
\hat{C}_{\mu\nu}^{(1)} &= - 2 H \hat{\mathfrak{D}}_{\mu\nu}^{(1)} - 2 H \left( g_{\mu\nu} + n u_\mu u_\nu \right) \mathfrak{T}^{(1)} \\
&\quad+ g_{\mu\nu} \frac{\kappa^2 \dot\phi}{(n-1) (n-2)} \mathfrak{S}^{(1)} - 2 u_\mu u_\nu (n-2-2\epsilon+2\delta) H^2 \frac{\mathfrak{S}^{(1)}}{\dot\phi} \eqend{.}
\end{split}
\end{align}
\end{subequations}
In the reverse direction, the linearised CDK tensors in terms of the FHH tensors are
\begin{subequations}
\label{eq:lin-cdk-from-fhh}
\begin{align}
\hat{\mathfrak{D}}_{\mu\nu}^{(1)} &= - \frac{1}{2 H} \left[ \hat{C}_{\mu\nu}^{(1)} - u_\mu u_\nu \left( u^\rho u^\sigma \hat{C}_{\rho\sigma}^{(1)} \right) - \frac{g_{\mu\nu} + u_\mu u_\nu}{n-1} \left( g^{\rho\sigma} + u^\rho u^\sigma \right) \hat{C}_{\rho\sigma}^{(1)} \right] \eqend{,} \\
\mathfrak{S}^{(1)} &= \frac{(n-1) \dot\phi \, g^{\mu\nu} \hat{C}_{\mu\nu}^{(1)}}{2 H^2 [ (n-2) (n-1-\epsilon) + 2 (n-1) \delta ]} \eqend{,} \\
\mathfrak{T}^{(1)} &= - \frac{1}{2 H (n-1)} \left( g^{\mu\nu} + u^\mu u^\nu \right) \hat{C}_{\mu\nu}^{(1)} + \frac{\kappa^2 \dot\phi}{2 H (n-1) (n-2)} \mathfrak{S}^{(1)} \eqend{,} \\
\begin{split}
\hat{\mathfrak{Z}}_{\mu\nu}^{(1)} &= \frac{1}{2} E_{\mu\nu}^{(1)} + (n-1) (n-2) H g_{\mu\nu} \mathfrak{T}^{(1)} \\
&\quad+ \frac{1}{2} u_\mu u_\nu \kappa^2 \dot\phi \mathfrak{S}^{(1)} + (n-2) \left( g_{\mu\nu} + u_\mu u_\nu \right) u^\rho \nabla_\rho \mathfrak{T}^{(1)} \eqend{,}
\end{split} \\
W_{\mu\nu\rho\sigma}^{(1)} &= C_{\mu\nu\rho\sigma}^{(1)} \eqend{,}
\end{align}
\end{subequations}
where we have made liberal use of the background relations~\eqref{background_phiv} to simplify the expressions.

\section{Discussion}
\label{sec:discussion}

We have reviewed two constructions of linear local gauge-invariant observables for cosmological perturbations on generic spatially flat single-field inflationary cosmologies. One stems from the work~\cite{fhh17}, the FHH observables~\eqref{eq:lin-fhh-def} and the other stems from the work~\cite{cdk17}, the linearisations of the CDK tensors~\eqref{eq:cdk-def}, which are locally and covariantly constructed from the metric and the inflaton field. We then gave an explicit and covariant transformation of the CDK tensors into the set~\eqref{eq:fhh-from-cdk}, the FHH tensors, together with the reverse transformation~\eqref{eq:cdk-from-fhh}. The FHH tensors are chosen such that their linearisations reproduce the FHH observables~\eqref{eq:lin-fhh-def}, thus closing the problem of identifying a Stewart--Walker type geometric interpretation for these observables (partially left open in~\cite{fhh17}). At the same time the linearisation of these transformations automatically gives explicit relations~\eqref{eq:lin-fhh-from-cdk} and~\eqref{eq:lin-cdk-from-fhh} between the CDK and FHH observables, thus showing explicitly the completeness of the CDK observables, which was proven in~\cite{fhh17} for the FHH ones but left open in~\cite{cdk17}.

It should be noted that the generic spatially flat inflationary cosmologies are just one of the special cases considered in~\cite{cdk17}. Other cases covered there include those with positive or negative spatial curvature, vanishing scalar potential, and also FLRW cosmologies without a scalar field. It would be interesting to perform an explicit analysis of the completeness of the corresponding linear local gauge-invariant observables on these backgrounds. Such an analysis would require a systematic approach to checking the completeness of sets of linear gauge-invariant observables on various backgrounds. Unfortunately, the method of~\cite{fhh17} do not easily generalise to other backgrounds, but some recent developments, to be reported elsewhere, may provide such an approach for a large class of spacetimes with Killing symmetries~\cite{aabk-private}.

\ack

This work is part of a project that has received funding from the European Union's Horizon 2020 research and innovation programme under the Marie Sk{\l}odowska-Curie grant agreement No. 702750 ``QLO-QG''.
IK was partially supported by the GA\v{C}R project 18-07776S and RVO: 67985840.

\providecommand\newblock{\ }
\bibliography{refs}

\end{document}